\documentclass[preprint,draft,prl,groupedaddress,onecolumn,10pt,tightenlines]{revtex4}

\input{tcilatex}

\begin{document}

\title{Extracting joint weak values with local, single-particle measurements}
\author{K.J. Resch$^{1}$ and A.M. Steinberg$^{1,2}$}
\affiliation{$^{1}$Institut f\"{u}r Experimentalphysik, Universit\"{a}t Wien \\
Boltzmanngasse 5, A-1090 Vienna, AUSTRIA\\
$^{2}$Department of Physics, University of Toronto, 60 St. George Street,
Toronto ON M5S 1A7, CANADA}

\begin{abstract}
Weak measurement is a new technique which allows one to describe the
evolution of postselected quantum systems. It appears to be useful for
resolving a variety of thorny quantum paradoxes, particularly when used to
study properties of pairs of particles. \ Unfortunately, such nonlocal or
joint observables often prove difficult to measure weakly in practice (for
instance, in optics -- a common testing ground for this technique -- strong
photon-photon interactions would be needed). Here we derive a general,
experimentally feasible, method for extracting these values from
correlations between single-particle observables.
\end{abstract}

\maketitle

In a ground-breaking work, Aharonov, Albert, and Vaidman (AAV) \cite{AAV}
introduced the concept of a ``weak measurement.'' In contrast to the usual
approach to quantum measurment, weak measurements can provide information
without greatly disturbing the evolution of a system. \ They are therefore
of special interest in the context of post-selected subensembles. \ In
particular, they have been used to ``resolve'' \cite{Hardyweak} Hardy's
Paradox \cite{Hardy}; to interpret and extend novel cavity-QED experiments %
\cite{orozco,wisemanQED}; to reconcile diverging views on which-path
experiments and uncertainty \cite{scully,walls,wisemanweak}; and to address
the long-standing controversy over tunneling times \cite{aeweak}. \ Often,
it is of interest to discuss the weak value of a nonlocal observable
(especially, of course, in proposals to study locality \cite%
{Hardy,Hardyweak,aetree}); unfortunately, it is usually difficult or
impossible to carry out such measurements \cite{Hardyweak,molmer}. In this
Letter, we demonstrate that the weak values of joint observables may be
readily extracted from experimentally realistic measurements.

In the usual quantum mechanical measurement process, correlations are
generated between the shift of the measurement pointer and the quantum
system through strong coupling interactions; this entanglement leads to
effective collapse. \ In AAV's weak measurement scheme, the coupling between
the measurement device and the system is extremely small and their
entanglement, and hence collapse, is avoided. \ The drawback is that on any
given event the resolution is insufficient to distinguish the different
eigenvalues of the observable. \ Nevertheless, by performing the experiment
many times on identically prepared systems one can reduce the measurement
uncertainty to any desired precision.

AAV considered the case where the weak measurement takes place between two
strong measurements called pre- and post-selection. \ Pre-selection of a
quantum state is commonplace in quantum mechanics experiments, and amounts
to the preparation of an initial quantum state, $\left| i\right\rangle .$ \
In a similar manner, post-selection projects out some final quantum state, $%
\left| f\right\rangle $. \ Only upon sucessful pre- and post-selection is
the state of the measurement apparatus considered. \ The weak measurement of
an obervable $A$ can take place via the same interaction Hamiltonian as a
strong (von Neumann) measurement,%
\begin{equation}
\mathcal{H}_{I}=g\hat{A}\hat{P_{x}},  \label{interaction}
\end{equation}%
where $\hat{A}$ is the hermitian operator corresponding to the observable $A$
of the quantum system, $g$ is a (real) coupling constant, and $\hat{P}_{x}$
is the measurement pointer momentum operator. \ This interaction shifts the
mean pointer position $\left\langle \hat{X}\right\rangle $ by an amount $%
\Delta x=K\left\langle \hat{A}\right\rangle \equiv gT\left\langle \hat{A}%
\right\rangle $, where $T$ is the length of time over which the interaction
is ``on''. \ The main result of AAV's seminal work is that for sufficiently
weak coupling strength $K$, the mean shift in the pointer position for a
pre- and post-selected subensemble is given by $\Delta x=K\func{Re}%
\left\langle \hat{A}\right\rangle _{W}$, where $\left\langle \hat{A}%
\right\rangle _{W}$ is the expression derived by AAV and termed by them the
``weak value'' \cite{AAV}:

\begin{equation}
\left\langle \hat{A}\right\rangle _{W}\equiv \frac{\left\langle f|\hat{A}%
|i\right\rangle }{\left\langle f|i\right\rangle }.  \label{weakvalue}
\end{equation}

The concept of weak measurement has been controversial since its inception %
\cite{commentsAAV}. \ In contrast to normal quantum mechanical expectation
values, weak values are not constrained to lie within the extremes of the
eigenvalue spectrum or to be real numbers. \ The controversy is only in the 
\emph{interpretation} of these strange results, since the predictions of the
formalism follow directly from quantum mechanics. To date, the \emph{%
predictions} of weak measurements have been tested only in optical
experiments. \ Ritchie et al. performed the first such test in an optical
Stern-Gerlach experiment \cite{weakexp}, based on some theoretical proposals %
\cite{AAV,duckproposal}. \ More recently, we have performed an experimental
realization of the Quantum Box Problem \cite{3Box,3boxexp}. \ In these
experiments, the single-particle observables of interest were coupled to the
transverse displacement of a light beam;\ this transverse position acted as
the measurement pointer.

Optical experiments are very limited in the kinds of interactions that are
available to experimentalists at the quantum level. \ One can realize an
effective interaction of the form of Eq. \ref{interaction}, for
single-photon polarization \cite{weakexp} or optical path measurements \cite%
{3boxexp}. \ However, there are more complicated types of observables for
which it is difficult to find a strong interaction Hamiltonian of the
desired form. \ An example of such an observable would be the joint
probability of one particle being found in region A and a second particle
being found in region B. Using AAV's approach, one would require a Kerr-like
nonlinearity at the single-photon level to couple the value of this joint
property into a shift of the measurement pointer. \ Such nonlinear effects
are very difficult to come by in optics \cite{kimble,franson,harris,switch},
and thus optical experiments might seem suited to only the simplest weak
measurement problems. \ More complicated theoretical proposals, such as the
ones involving Hardy's Paradox \cite{Hardyweak,Hardy}, nonlocality of single
particle \cite{aetree}, and extensions of the Quantum Box Problem \cite%
{aewheeler} are extremely difficult to implement if they require optical
nonlinearities at the quantum level, or other multi-particle or nonlocal
interactions. One theoretical proposal has been made for carrying out such
joint weak measurements with trapped ions, relying on the natural, strong,
interparticle interaction \cite{molmer}; to date, no such quantitites have
been measured in the laboratory.

In this work, we show that it is experimentally possible to measure these
complicated \emph{joint weak values} using local, single-particle
interaction Hamiltonians by studying the correlations between two different
pointer variables (either two different one-dimensional pointers or a single
two-dimensional pointer). \ We derive an explicit formula for finding the
weak value of an observable which is the product of two operators; this is
sufficient to test all of the aforementioned theoretical proposals in
optical systems. \ (The theoretical approach is general enough that it can
be extended to products of N operators by using N pointer dimensions.)

We begin by rederiving AAV's result for the deflection of a measurement
pointer through the weak measurement procedure using the interaction
Hamiltonian from Eq. \ref{interaction}. \ Throughout our calculations we
assume that the operators $\hat{A}$ and $\hat{P}_{x}$ commute, which is a
very natural assumption since one operator corresponds to a property of the
quantum system and the other to the measurement apparatus.

The quantum system is initially prepared in the state $\left| i\right\rangle 
$. \ We consider a measurement pointer initially prepared in a gaussian
state centered at zero: 
\begin{equation}
\phi (x)=\left( \frac{1}{\sqrt{2\pi }\sigma }\right) ^{\frac{1}{2}}\exp
\left( -\frac{x^{2}}{4\sigma ^{2}}\right) ,
\end{equation}%
where $\sigma $ is the rms\ width of the probability distribution. \ We now
calculate the expected displacement of the measurement pointer when the
quantum state is post-selected in the final state $\left| f\right\rangle $,
using the rule 
\begin{equation}
\left\langle \hat{X}\right\rangle _{\mathrm{fi}}=\frac{\left\langle \hat{P}%
(f)\hat{X}\right\rangle }{\left\langle \hat{P}(f)\right\rangle }\;,
\end{equation}%
where $\left\langle \hat{X}\right\rangle _{\mathrm{fi}}$ is the expectation
value of $\hat{X}$ given a successful post-selection of the final state $%
\left| f\right\rangle ,$ and $\hat{P}(f)$ is the projector $\left|
f\right\rangle \left\langle f\right| $ onto the final state \cite{aeweak}.
This amounts to finding the expectation value of the operator $\widehat{O}%
_{x}\equiv \left| f\right\rangle \left\langle f\right| \hat{X}$ and
renormalizing. \ We employ the Heisenberg equation of motion to propagate
the expectation value of $\widehat{O}_{x}$ forward in time to first order: \
\qquad \qquad 
\begin{equation}
\left\langle \widehat{O}_{x}(t)\right\rangle \simeq \left\langle \widehat{O}%
_{x}(0)\right\rangle +\frac{it}{\hbar }\left\langle \left[ \mathcal{H},%
\widehat{O}_{x}\right] \right\rangle .
\end{equation}%
This first-order approximation is only valid for weak values of the coupling 
$K$. Our measurement pointer was originally centred at zero, and therefore
the first term on the right-hand side vanishes. \ The expectation value of $%
\widehat{O}_{x}$ to first-order is\ 
\begin{eqnarray}
\left\langle \widehat{O}_{x}(t)\right\rangle &\simeq &\frac{it}{\hbar }%
g\left\langle \phi \left| \left\langle i\left| \left\{ \hat{A}\hat{P}%
_{x}\left| f\right\rangle \left\langle f\right| \hat{X}-\left|
f\right\rangle \left\langle f\right| \hat{X}\hat{A}\hat{P}_{x}\right\}
\right| i\right\rangle \right| \phi \right\rangle \\
&=&\frac{it}{\hbar }g\left[ \left\langle i\left| \hat{A}\right|
f\right\rangle \left\langle f|i\right\rangle \left\langle \phi \left| \hat{P}%
_{x}\hat{X}\right| \phi \right\rangle -\left\langle i|f\right\rangle
\left\langle f\left| \hat{A}\right| i\right\rangle \left\langle \phi \left| 
\hat{X}\hat{P}_{x}\right| \phi \right\rangle \right] .
\end{eqnarray}%
The values of the pointer integrals are $\left\langle \phi \left| \hat{P}_{x}%
\hat{X}\right| \phi \right\rangle =-i\hbar /2$ and $\left\langle \phi \left| 
\hat{X}\hat{P}_{x}\right| \phi \right\rangle =i\hbar /2$. \ We are justified
in truncating the time-dependence of the system at first order only if the
higher-order terms are small. \ Second-order terms will contain integrals
with quadratic terms in the momentum and linear in position and are
therefore zero by symmetry; however third-order terms will be proportional
to $1/\sigma ^{2}$. \ These terms will contribute negligibly to the
expectation value of the weak value provided $Ka\ll \sigma $, where $a$ is
the characteristic scale of $\hat{A}$; that is to say that the uncertainty
in the pointer position is much larger than the strength of the interaction
-- this was AAV's condition for weakness. \ We now renormalize to obtain the
conditional expectation value of $\hat{X}$ given pre-selection and
postselection, 
\begin{equation}
\left\langle \hat{X}\right\rangle _{\mathrm{fi}}=\frac{K}{2}\left[ \frac{%
\left\langle i\left| \hat{A}\right| f\right\rangle }{\left\langle
i|f\right\rangle }+\frac{\left\langle f\left| \hat{A}\right| i\right\rangle 
}{\left\langle f|i\right\rangle }\right] .  \label{resingle}
\end{equation}%
Thus we see that the shift of the pointer, conditioned on appropriate
postselection is indeed proportional to the weak value $\left\langle \hat{A}%
\right\rangle _{W}$ from Eq. \ref{weakvalue}: 
\begin{equation}
\func{Re}\left\langle \hat{A}\right\rangle _{W}=\frac{\left\langle \hat{X}%
\right\rangle _{\mathrm{fi}}}{K}.  \label{resingle2}
\end{equation}

It is well known that weak values are not constrained to be real numbers. \
The imaginary part of the weak value yields a shift in the variable
conjugate to the pointer position (i.e. $\hat{P_{x}}$), related to the
back-action of the measurement \cite{aeweak}. \ Using the same initial state
of the measurement apparatus, one can go through a similar calculation for
the expectation value of the operator $\widehat{O}_{p}\equiv \left|
f\right\rangle \left\langle f\right| \hat{P}_{x}.$ \ The result of this
calculation for the imaginary part of the weak value is: 
\begin{equation}
\func{Im}\left\langle \hat{A}\right\rangle _{W}=\frac{2\sigma ^{2}}{\hbar }%
\frac{\left\langle \hat{P}_{x}\right\rangle _{\mathrm{fi}}}{K}.
\label{imsingle}
\end{equation}%
Note that, unlike the real part, the imaginary part of the weak value
depends quadratically on the width of the pointer distribution, $\sigma .$ \
This means that for more and more ideal weak measurements (i.e. as $\sigma
\rightarrow \infty $) the actual $\emph{physical}$ shift in the pointer
momentum, $\left\langle \hat{P}_{x}\right\rangle _{\mathrm{fi}},$ will tend
to zero. \ 

We now proceed to calculate an expression for joint weak values. \ Instead
of employing a nonlinear optical Hamiltonian to measure a joint weak value $%
\left\langle \hat{A}\hat{B}\right\rangle _{W}$ of two observables $A$ and $B$
(with corresponding hermitian operators $\hat{A}$ and $\hat{B}$), we use the
local single-particle interaction Hamiltonian%
\begin{equation}
\mathcal{H}=g_{A}\hat{A}\hat{P}_{x}+g_{B}\hat{B}\hat{P}_{y},
\label{practicalHam}
\end{equation}%
where $g_{A}$ and $g_{B}$ are coupling constants and $\hat{P}_{x}$ and $\hat{%
P}_{y}$ are the momentum operators conjugate to two commuting pointer
operators $\hat{X}$ and $\hat{Y}$. Our system is prepared in the quantum
state $\left| i\right\rangle $ and the measurement apparatus in a state $%
\left| \Phi \right\rangle \equiv \left| \phi _{x}\right\rangle \left| \phi
_{y}\right\rangle $ described by a two-dimensional gaussian wavepacket, 
\begin{equation}
\Phi (x,y)=\frac{1}{\sqrt{2\pi \sigma _{x}\sigma _{y}}}\exp \left( -\frac{%
x^{2}}{4\sigma _{x}^{2}}\right) \exp \left( -\frac{y^{2}}{4\sigma _{y}^{2}}%
\right) ,  \label{2dmeasinitial}
\end{equation}%
where $\sigma _{x,y}$ are the rms\ widths of the probability distribution in 
$x$ and $y$, respectively.

In this case, we are interested in the joint weak value $\left\langle \hat{A}%
\hat{B}\right\rangle _{W}$, and therefore need an expression for a pointer
deflection that contains contributions from deflections in the x-direction
and in the y-direction after the successful post-selection of the quantum
system into the state $\left| f\right\rangle $. \ By analogy with the
previous calculation, we find the expectation value of the operator, $%
\widehat{O}_{xy}\equiv \left| f\right\rangle \left\langle f\right| \hat{X}%
\hat{Y}$. \ We use the Heisenberg equation of motion recursively to expand
the right-hand side to \emph{second} order: 
\begin{equation}
\left\langle \widehat{O}_{xy}(t)\right\rangle \simeq \left\langle \widehat{O}%
_{xy}(0)\right\rangle +i\frac{t}{\hbar }\left\langle \left[ \mathcal{H},%
\widehat{O}_{xy}\right] \right\rangle -\frac{1}{2!}\left( \frac{t^{2}}{\hbar
^{2}}\right) \left\langle \left[ \mathcal{H},\left[ \mathcal{H},\widehat{O}%
_{xy}\right] \right] \right\rangle .
\end{equation}%
The zeroth-order term, $\left\langle \widehat{O}_{xy}(0)\right\rangle ,$ in
the expectation value is zero since $\left| \Phi \right\rangle $ is even
with respect to both $x$ and $y$, while $\widehat{O}$ is odd in both. The
first-order commutator is given by: 
\begin{equation}
\left[ \mathcal{H},\widehat{O}_{xy}\right] =\left( g_{A}\hat{A}\hat{P}%
_{x}+g_{B}\hat{B}\hat{P}_{y}\right) \left| f\right\rangle \left\langle
f\right| \hat{X}\hat{Y}-\left| f\right\rangle \left\langle f\right| \hat{X}%
\hat{Y}\left( g_{A}\hat{A}\hat{P}_{x}+g_{B}\hat{B}\hat{P}_{y}\right) .
\end{equation}%
Each term in this expression is even with respect to either $x$ or $y$ but
not both, and therefore has a vanishing expectation value for the pointer
state $\left| \Phi \right\rangle $. Therefore, to first order, since the
measurement apparatus can receive a kick in either the $x$-direction or the $%
y$-direction but never both, there is no shift in the expectation value of $%
\widehat{O}$. We require the second-order term, where kicks in both $x$ and $%
y$ can appear, to find a nonzero value. \ The second-order commutator
contains 16 terms of which only 8 contain the product of $\hat{P}_{x}$ and $%
\hat{P}_{y}$ and can therefore have nonzero expectation values. \ These
remaining 8 terms can be simplified by collecting 2 identical pairs of terms
to%
\begin{equation}
\left\langle \left[ \mathcal{H},\left[ \mathcal{H},\widehat{O}_{xy}\right] %
\right] \right\rangle =g_{A}g_{B}\left\langle 
\begin{array}{c}
\hat{B}\hat{A}\hat{P}_{x}\hat{P}_{y}\widehat{O}_{xy}+\hat{A}\hat{B}\hat{P}%
_{x}\hat{P}_{y}\widehat{O}_{xy}-2\hat{A}\hat{P}_{x}\widehat{O}_{xy}\hat{B}%
\hat{P}_{y} \\ 
-2\hat{B}\hat{P}_{y}\widehat{O}_{xy}\hat{A}\hat{P}_{x}+\widehat{O}_{xy}%
\hat{A}\hat{B}\hat{P}_{x}\hat{P}_{y}+\widehat{O}_{xy}\hat{B}\hat{A}\hat{P}%
_{x}\hat{P}_{y}%
\end{array}%
\right\rangle .
\end{equation}%
We further simplify this result by evaluating the integrals for the
measurement apparatus, $\left\langle \phi \left| \hat{P}_{x}\hat{X}\right|
\phi \right\rangle =-i\hbar /2$ and $\left\langle \phi \left| \hat{X}\hat{P}%
_{x}\right| \phi \right\rangle =i\hbar /2$ . Thus,%
\begin{eqnarray}
\left\langle \widehat{O}_{xy}(t)\right\rangle &=&-\frac{1}{2!}\left( \frac{%
t^{2}}{\hbar ^{2}}\right) g_{A}g_{B}\left[ 
\begin{array}{c}
-\frac{\hbar ^{2}}{4}\left\langle i\left| \hat{B}\hat{A}\right|
f\right\rangle \left\langle f|i\right\rangle -\frac{\hbar ^{2}}{4}%
\left\langle i\left| \hat{A}\hat{B}\right| f\right\rangle \left\langle
f|i\right\rangle \\ 
-\frac{\hbar ^{2}}{2}\left\langle i\left| \hat{A}\right| f\right\rangle
\left\langle f\left| \hat{B}\right| i\right\rangle -\frac{\hbar ^{2}}{2}%
\left\langle i\left| \hat{B}\right| f\right\rangle \left\langle f\left| 
\hat{A}\right| i\right\rangle \\ 
-\frac{\hbar ^{2}}{4}\left\langle i|f\right\rangle \left\langle f\left| 
\hat{A}\hat{B}\right| i\right\rangle -\frac{\hbar ^{2}}{4}\left\langle
i|f\right\rangle \left\langle f\left| \hat{B}\hat{A}\right| i\right\rangle%
\end{array}%
\right] \\
&=&\frac{1}{2}K_{x}K_{y}\func{Re}\left[ \left\langle i|f\right\rangle
\left\langle f\left| \frac{\hat{A}\hat{B}+\hat{B}\hat{A}}{2}\right|
i\right\rangle +\left\langle i\left| \hat{A}\right| f\right\rangle
\left\langle f\left| \hat{B}\right| i\right\rangle \right] \;,
\end{eqnarray}%
where $K_{x,y}\equiv g_{A,B}t$ in analogy with our one-dimensional
definition. We again employ the notation $\left\langle \hat{X}\hat{Y}%
\right\rangle _{\mathrm{fi}}$ to mean the expectation value for $\hat{X}%
\hat{Y}$ given a successful post-selection. \ This conditional expectation
value requires division by the success probability for normalization such
that%
\begin{eqnarray}
\left\langle \hat{X}\hat{Y}\right\rangle _{\mathrm{fi}} &=&\frac{1}{2}%
K_{x}K_{y}\frac{\func{Re}\left[ \left\langle i|f\right\rangle \left\langle
f\left| \frac{\hat{A}\hat{B}+\hat{B}\hat{A}}{2}\right| i\right\rangle
+\left( \left\langle i\left| \hat{A}\right| f\right\rangle \left\langle
f\left| \hat{B}\right| i\right\rangle \right) \right] }{\left| \left\langle
i|f\right\rangle \right| ^{2}} \\
&=&\frac{1}{2}K_{x}K_{y}\func{Re}\left[ \frac{\left\langle f\left| \frac{%
\hat{A}\hat{B}+\hat{B}\hat{A}}{2}\right| i\right\rangle }{\left\langle
f|i\right\rangle }+\left\langle \hat{A}\right\rangle _{W}^{\ast
}\left\langle \hat{B}\right\rangle _{W}\right] ,  \label{realweakAB}
\end{eqnarray}%
where $A_{W}$ and $B_{W}$ are the weak values of the individual operators as
defined by AAV (Eq. \ref{weakvalue}), and can be extracted from experimental
measurement as shown in the previous theory section. \ We again are making
an approximation when truncating our evolution at second order. \ Third
order terms are identically zero due to symmetry; however fourth-order terms
are not. \ Similar to the single weak values, these higher order terms
contribute negligibly provided that $K_{x,y}$ are sufficiently small with
respect to the widths $\sigma _{x,y}$, in direct analog with the condition
we imposed in the one-dimensional case. We rearrange Eq. \ref{realweakAB} to
obtain an expression for the weak value of $(\hat{A}\hat{B}+\hat{B}\hat{A}%
)/2,$ 
\begin{equation}
\func{Re}\left\langle \frac{\hat{A}\hat{B}+\hat{B}\hat{A}}{2}\right\rangle
_{W}=2\frac{\left\langle \hat{X}\hat{Y}\right\rangle _{\mathrm{fi}}}{%
K_{x}K_{y}}-\func{Re}\left( \left\langle \hat{A}\right\rangle _{W}^{\ast
}\left\langle \hat{B}\right\rangle _{W}\right) ,  \label{symmresult}
\end{equation}%
where the subscript $W$ refers to the weak value. \ If $\hat{A}$ and $\hat{B}
$ are commuting observables (as must be the case, for instance, if they
operate on disjoint subsystems), then $\hat{A}\hat{B}$ is Hermitian and the
left-hand side is equal to $\func{Re}(\hat{A}\hat{B})_{W}$, the desired
result: 
\begin{equation}
\func{Re}\left\langle \hat{A}\hat{B}\right\rangle _{W}=2\frac{\left\langle 
\hat{X}\hat{Y}\right\rangle _{\mathrm{fi}}}{K_{x}K_{y}}-\func{Re}\left(
\left\langle \hat{A}\right\rangle _{W}^{\ast }\left\langle \hat{B}%
\right\rangle _{W}\right) .  \label{result}
\end{equation}%
For $\hat{A}\hat{B}$ non-Hermitian, the AAV coupling Hamiltonian $\mathcal{H}%
=g\hat{A}\hat{B}\hat{P}$ would be unphysical, and only the symmetric form of
Eq. \ref{symmresult} could correspond to an observable (assuming $\hat{A}$
and $\hat{B}$ Hermitian and $g$ real). Note that other operator orderings
have also been studied in certain contexts \cite{muga}.

Through an analogous calculation we can obtain an expression for the
imaginary part of the weak value of the operator $\hat{A}\hat{B}$ by
calculating the first nonzero term in the expectation value of the operator $%
\widehat{O}_{xp_{y}}\equiv \left| f\right\rangle \left\langle f\right| \hat{X%
}\hat{P}_{y}$ under the same interaction Hamiltonian from Eq. \ref%
{practicalHam}. \ The imaginary part of the weak value is:%
\begin{equation}
\func{Im}\left\langle \frac{\hat{A}\hat{B}+\hat{B}\hat{A}}{2}\right\rangle
_{W}=\frac{4\sigma _{y}^{2}}{\hbar }\frac{\left\langle \hat{X}\hat{P}%
_{y}\right\rangle _{\mathrm{fi}}}{K_{x}K_{y}}-\func{Im}\left( \left\langle 
\hat{A}\right\rangle _{W}^{\ast }\left\langle \hat{B}\right\rangle
_{W}\right)  \label{result2}
\end{equation}%
which can again be further simplified on the left-hand side if $\hat{A}$ and 
$\hat{B}$ commute. \ As we noted in the single weak value case, the physical
implication (i.e. the shift in $\left\langle \hat{X}\hat{P}_{y}\right\rangle
_{\mathrm{fi}}$) of the imaginary part of the weak value becomes smaller and
smaller in the limit as $\sigma \rightarrow \infty $. \ 

Note that all of the quantities in Eqs. \ref{result} and \ref{result2} are
experimentally measurable in the same experiment. \ The quantities $%
\left\langle \hat{X}\hat{Y}\right\rangle _{\mathrm{fi}}$ and $\left\langle 
\hat{X}\hat{P}_{y}\right\rangle _{\mathrm{fi}}$ are extracted from the
two-dimensional pointer deflections, and $\left\langle \hat{A}\right\rangle
_{W}^{\ast }$ and $\left\langle \hat{B}\right\rangle _{W}$ can be measured
using the method from Eq. \ref{resingle2} and Eq. \ref{imsingle}.

Since AAV's original work on weak measurement, their predictions have been
demonstrated experimentally in several optical experiments. \ In a number of
circumstances, it has become clear that weak measurements of nonlocal and/or
multi-particle observables should be of great interest. \ The interaction
Hamiltonians required to accomplish such measurements using the original AAV
approach are often experimentally inaccessible, e.g., demanding strong
interactions between individual photons. \ We have developed an alternative
method for extracting the desired joint weak values that can be implemented
using only local, single-particle Hamiltonians. \ In addition to being
necessary for the resolution of certain quantum paradoxes \cite%
{Hardyweak,3boxexp,aetree}, the ability to discuss the correlations present
in postselected systems should prove indispensible for investigating the
role of postselection in quantum information \cite{klm}. \ The new approach
described here will be essential for the broad experimental application of
joint weak measurements.

We would like to thank Jeff Lundeen and Morgan Mitchell for helpful
discussions. \ This work was funded by the U.S. Air Force Office of
Scientific Research (F49620-01-1-0468), Photonics Research Ontario, and
NSERC. \

\end{document}